\documentclass[runningheads]{llncs}
\usepackage{graphicx}
\usepackage{url}

\usepackage[sort&compress]{natbib}

\newcommand{\journal}[1]{#1}
\newcommand{\conferenceonly}[1]{\relax}
\newtheorem{thm}{Theorem}

\newcommand{\figname}{Figure.}

\begin{document}

\title{Delta-confluent Drawings
\thanks{%
Work by the first author is supported by NSF grant CCR-9912338. 
Work by the second and the third author is supported by 
NSF grants CCR-0098068, CCR-0225642, and DUE-0231467.}%
}

\titlerunning{Delta-confluent Drawings}

\author{David~Eppstein%
\and Michael~T.~Goodrich%
\and Jeremy~Yu~Meng%
}

\authorrunning{Eppstein, Goodrich, and Meng}


\institute{
School of Information and Computer Science,\\
University of California, Irvine,\\
Irvine, CA 92697, USA\\
\email{\{eppstein, goodrich, ymeng\}@ics.uci.edu}
}

\maketitle

\begin{abstract}
We generalize the \emph{tree-confluent} graphs to a 
broader class of graphs called \emph{$\Delta$-confluent} graphs. 
This class of graphs and 
distance-hereditary graphs,
a well-known class of graphs, 
coincide.
Some results about the visualization of $\Delta$-confluent graphs 
are also given. 
\end{abstract}

\section{Introduction}
\label{sec:intro}

\begin{figure}[htb]
  \centering
  \includegraphics[width=0.6\textwidth]{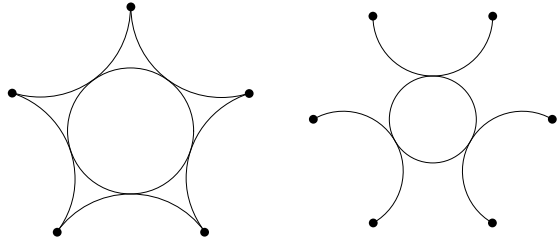}
  \caption{Confluent drawings of $K_5$ and $K_{3,3}$.}
  \label{fig:tra-cir}
\end{figure}

\emph{Confluent Drawing} is an approach to visualize non-planar
graphs in a planar way~\cite{degm-cdvnd-04}. 
The idea is simple:  
we allow groups of edges 
to be merged together and 
drawn as tracks (similar to train tracks).  
This method allows us to draw, 
in a crossing-free manner, graphs that would 
have many crossings in their normal drawings. 
Two examples are shown in \figname~\ref{fig:tra-cir}. 
In a confluent
drawing, two nodes are connected if and only if 
there is a smooth curve path 
from one to the other 
without making sharp turns or double backs,
although multiple realizations of a graph edge 
in the drawing is allowed. 

More formally, 
a curve is \emph{locally-monotone} if it contains no
self intersections and no
sharp turns, that is, it contains no 
point with left and right tangents
that form an angle less than or equal to $90$ degrees.
Intuitively, a locally-monotone curve is like a single train track, which
can make no sharp turns.
Confluent drawings are
a way to draw graphs in a planar manner by
merging edges together into \emph{tracks}, which are the unions of 
locally-monotone curves.

An undirected graph $G$ is \textit{confluent} if and only if there exists a
drawing $A$ such that:

\vspace*{-8pt}
\begin{itemize}
\setlength{\itemsep}{0pt}
\setlength{\parsep}{0pt}
\item 
There is a one-to-one mapping between the vertices in $G$ and
  $A$, so that, for each vertex $v \in V(G)$, there is a corresponding vertex 
  $v' \in A$, which has a unique point placement in the plane.
\item 
  There is an edge $(v_i,v_j)$ in $E(G)$
  if and only if there is a locally-monotone curve $e'$
  connecting $v_i'$ and $v_j'$ in $A$.  
\item 
$A$ is planar.
That is, while locally-monotone curves in $A$ can share overlapping portions,
no two can cross.
\end{itemize}
\vspace*{-8pt}

We assume readers have basic knowledge about graph 
theory and we will use conventional terms and notations 
of graph theory without defining them.
All graphs considered in this paper are simple graphs,
i.e., without loop or multi-edge.
Confluent graphs are closely related to planar graphs.
It is, however, very hard to check whether a given graph 
can be drawn confluently.  
The complexity of recognizing confluent graphs is still 
open and the problem is expected to be hard. 
Hui, Schaefer and~{\v S}tefankovi{\v c}~\cite{hss-ttcd-05}
define the notion of \emph{strong confluency} and show that 
strong confluency can be recognized in \textbf{NP}.  
It is then of interest to study
classes of graphs that can or can not be drawn confluently.
Several classes of confluent graphs,
as well as several classes of non-confluent graphs, 
have been listed~\cite{degm-cdvnd-04}. 

In this paper we continue in the positive direction
of this route. 
We describe $\Delta$-confluent graphs, 
a generalization of \emph{tree-confluent} 
graphs~\cite{hss-ttcd-05}.  We discuss problems of 
embedding trees with internal degree three,
including embeddings on the hexagonal grid, 
which is related to $\Delta$-confluent drawings with 
large angular resolution, and show that 
$O(n\log n)$ area is enough for a $\Delta$-confluent drawing 
of a $\Delta$-confluent graph with $n$ vertices on the hexagonal grid.

Note that although the method of merging groups of edges is also 
used to reduce crossings 
in \emph{confluent layered drawings}~\cite{egm-cld-05}, 
edge crossings are allowed to exist in a confluent 
layered drawing.

\section{$\Delta$-confluent graphs}
\label{sec:delta-confl-graphs}

Hui, Schaefer and~{\v S}tefankovi{\v c}~\cite{hss-ttcd-05} 
introduce the idea of \emph{tree-confluent} graphs.  
A graph is \emph{tree-confluent} if and only if it is 
represented by a planar train track system which is 
topologically a tree.
It is also shown in their paper that 
the class of tree-confluent graphs 
are equivalent to the class of
chordal bipartite graphs.

The class of tree-confluent graphs can be extended 
into a wider class of graphs if we allow 
one more powerful type of junctions.

A \emph{$\Delta$-junction} is a 
structure 
where three paths 
are allowed to meet in a three-way complete junction.
The connecting point is call a \emph{port} of the junction.
A \emph{$\Lambda$-junction} is a broken $\Delta$-junction 
where two of the three ports are disconnected 
from each other (exactly same as the \emph{track} 
defined in
the tree-confluent drawing~\cite{hss-ttcd-05}). 
The two disconnected paths are called \emph{tails} 
of the $\Lambda$-junction and the remaining one is 
called \emph{head}.

\begin{figure}[htb]
  \centering
  \includegraphics[width=0.6\textwidth]{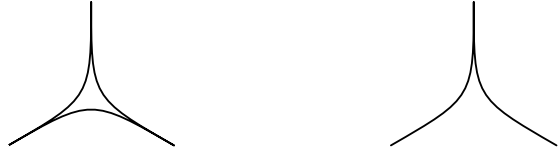}
  \caption{$\Delta$-junction and $\Lambda$-junction.}
  \label{fig:tracks}
\end{figure}

A \emph{$\Delta$-confluent drawing} is 
a confluent drawing in which 
every junction in the drawing 
is either a $\Delta$-junction,
or a $\Lambda$-junction,
and 
if we replce every junction in the drawing with 
a new vertex, we get a tree.
A graph $G$ is $\Delta$-confluent 
if and only if it has a $\Delta$-confluent drawing.  

The class of cographs in~\cite{degm-cdvnd-04} and the class of
tree-confluent graphs in~\cite{hss-ttcd-05} are both included 
in the class of $\Delta$-confluent graphs.
We observe that the class of $\Delta$-confluent graphs are 
equivalent to the class of distance-hereditary graphs.

\subsection{Distance-hereditary graphs}
\label{sec:dist-hered-graphs}

A \emph{distance-hereditary} graph is a connected graph 
in which every induced path is isometric. 
That is, the distance of any two vertices in an induced path 
equals their distance in the graph~\cite{bm-dhg-86}. 
Other characterizations have been found for distance-hereditary
graphs: forbidden subgraphs, properties of cycles, etc.
Among them, the following one is most interesting to us:
\begin{thm}
\emph{\cite{bm-dhg-86}} Let $G$ be a finite graph 
with at least two vertices.
Then $G$ is distance-hereditary if and only if
$G$ is obtained from $K_2$ by a sequence of 
one-vertex extensions: attaching pendant vertices
and splitting vertices.
\end{thm}
Here attaching a pendant vertex to $x$ means 
adding a new vertex $x'$ to $G$ and making it 
adjacent to $x$ so $x'$ has degree one;
and splitting $x$ means adding a new vertex $x'$ to $G$ 
and making it adjacent to either $x$ and all neighbors of $x$,
or just all neighbors of $x$.
Vertices $x$ and~$x'$ forming a split pair are called 
\emph{true twins} (or \emph{strong siblings})
if they are adjacent, 
or \emph{false twins} (or \emph{weak siblings}) otherwise.

By reversing the above extension procedure, 
every finite distance-hereditary graph $G$ can be 
reduced to $K_2$
in a sequence of one-vertex operations: 
either delete a pendant vertex 
or identify a pair of twins $x'$ and~$x$.
Such a sequence is called an 
\emph{elimination sequence} (or a \emph{pruning sequence}).

\begin{figure}[htb]
  \centering
  \includegraphics[width=.5\textwidth]{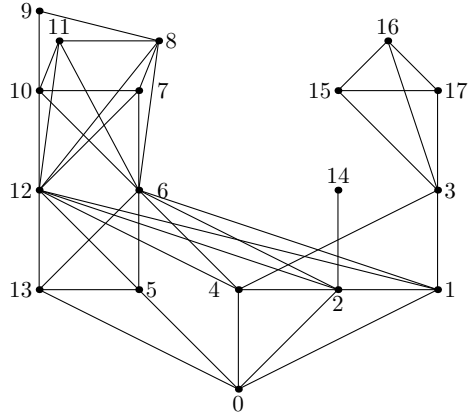}
  \caption{A distance-hereditary graph $G$}
  \label{fig:dhex1}
\end{figure}

In the example distance-hereditary graph $G$ 
of \figname~\ref{fig:dhex1}, 
the vertices are labelled reversely according to 
an elimination sequence of $G$:

$17$ merged into $16$,
$16$ merged into $15$,
$15$ cut from $3$,
$14$ cut from $2$,
$13$ merged into $5$,
$12$ merged into $6$,
$10$ merged into $8$,
$11$ merged into $7$,
$9$ cut from $8$,
$8$ merged into $7$,
$7$ cut from $6$,
$6$ merged into $0$,
$5$ cut from $0$,
$4$ merged into $1$,
$3$ cut from $1$,
$2$ merged into $1$.

The following theorem 
states that the class of distance hereditary graphs and 
the class of $\Delta$-confluent graphs are equivalent.
\begin{thm}
  A graph $G$ is distance hereditary if and only if 
it is $\Delta$-confluent.
\end{thm}
\emph{Proof.}  Assume $G$ is distance hereditary. 
We can compute the elimination sequence of $G$, 
then apply 
an algorithm, which will be described in Section~\ref{sec:elim-sequ-delta},
to get a $\Delta$-confluent drawing of $G$.  Thus $G$ is 
$\Delta$-confluent. 

On the other hand, given a $\Delta$-confluent graph~$G$
in form of its $\Delta$-confluent drawing $A$,
we can apply the following operations on the drawing~$A$:
\begin{enumerate}
\item \emph{contraction}. 
  If two vertices $y$ and~$y'$ in $A$
  are connected to two ports of a $\Delta$-junction, 
  or $y$ and~$y'$ are connected to the two tails
  of a $\Lambda$-junction respectively, 
  then contract $y$ and~$y'$ into a 
  new single vertex, 
  and replace the junction with this new vertex.
\item \emph{deletion}. 
  If two vertices $y$ and~$y'$ in $A$ 
  are connected by a $\Lambda$-junction,
  $y$ is connected to the head and $y'$ to one tail,
  remove $y'$ and replace the junction with~$y$.
\end{enumerate}

%
%
It is easy to observe that contraction in the drawing~$A$ corresponds
to identifying a pair of twins in~$G$; and deletion corresponds to 
removing a pendant vertex in~$G$.

It is always possible to apply an operation on two vertices
connected by a junction because the underlying graph is a tree. 
During each operation one junction is replaced.  
Since the drawing is finite, the number of junctions is finite. 
Therefore, we will reach a point at which 
the last junction is replaced. 
After that the drawing reduces to a pair of 
vertices connected by an edge,
and the corresponding~$G$ reduces to a $K_2$.
Therefore $G$ is a distance-hereditary graph.
 
This completes the proof of the equivalence between 
$\Delta$-confluent graphs and distance-hereditary graphs. \qed

\subsection{Elimination Sequence to $\Delta$-confluent tree}
\label{sec:elim-sequ-delta}

The recognition problem of distance-hereditary graphs 
is solvable in linear time (see~\cite{bm-dhg-86,hm-csg-90}).
The elimination sequence (ordering) can also be computed in 
linear time.  Using the method of, for example,
Damiand et al.~\cite{dhp-spgra-01}
we can obtain an elimination sequence $L$ for $G$ of 
\figname~\ref{fig:dhex1}:

By using the elimination sequence reversely, 
we construct a tree structure of 
the $\Delta$-confluent drawing of $G$.  This tree structure has 
$n$ leaves and $n-1$ internal nodes.  
Every internal node has 
degree of three.  
The internal nodes represent our $\Delta$- and $\Lambda$-junctions.  
The construction is as follows.

\begin{itemize}
\item While $L$ is non-empty do:
  \begin{itemize}
  \item Get the last $item$ from $L$
  \item If $item$ is ``$b$ merged into $a$''
    \begin{itemize}
    \item If edge $(a,b)\in E(G)$, then replace $a$ with a $\Delta$ 
      conjunction using any of its three connectors, 
      connect $a$ and $b$ to the other two connectors
      of the $\Delta$ conjunction; otherwise replace $a$ with a 
      $\Lambda$ conjunction using its head and 
      connect $a$ and $b$ to its two tails. 
    \end{itemize}
  \item Otherwise $item$ is ``$b$ cut from $a$'', 
    replace $a$ with a $\Lambda$
    conjunction using one of its tails, connect $a$ to the head and $b$ 
    to the other tail left.
  \end{itemize}
\end{itemize}

Clearly the structure we obtain is 
indeed a tree.
Once the tree structure is constructed, 
the $\Delta$-confluent drawing can be computed by visualizing 
this tree structure with its internal nodes replaced by 
$\Delta$- and $\Lambda$-junctions.

\section{Visualizing the $\Delta$-confluent graphs}
\label{sec:visu-delta-confl}

There are many methods to visualize the underlying 
topological tree of a $\Delta$-confluent drawing.  
Algorithms for drawing trees have been studied extensively
(see~\cite{%
bk-abtl-80,%
cdp-noaau-92,%
e-dft-92,%
ell-ttdc-93,%
g-ulebt-89,%
ggt-aeutd-93,%
i-dtg-90,%
mps-phvdb-94,%
rt-tdt-81,%
sr-cdtn-83,%
t-tda-81,%
w-npagt-90,%
ws-tdt-79%
} for examples).
Theoretically all the tree visualization methods 
can be used to lay out the underlying tree of 
a $\Delta$-confluent drawing,
although free tree drawing techniques 
might be more suitable.
We choose the following two tree drawing approaches
that both yield large angular resolution ($\ge \pi/2$), 
because in drawings with large angular resolution,
each junction
lies in a center-like position among the nodes
connected to it,
so junctions
are easy to perceive and 
paths are easy to follow. 

\subsection{Orthogonal straight-line $\Delta$-confluent drawings}
\label{sec:orth-stra-line}

The first tree drawing method is 
the orthogonal straight-line 
tree drawing method.  
In the drawings by this method,
every edge is drawn as a straight-line segment and every 
node is drawn at a grid position. 

Pick an arbitrary leaf node $l$ 
of the underlying tree
as root and make this free tree a rooted tree $T$
(alternatively one can adopt the elimination 
hierarchy tree of a distance-hereditary graph for use here.)
It is easy to see that $T$ is a binary tree
because every internal node of the underlying tree has degree three. 
We can then apply any known orthogonal 
straight-line drawing algorithm for trees 
(\citep[e.g.][]{l-aeglv-80,l-aeglv-83,%
v-ucvc-81,bk-abtl-80,cgkt-oaars-97,c-osldt-99})
on $T$ to obtain a layout.
After that, replace drawings of internal nodes 
with their corresponding junction drawings. 

\subsection{Hexagonal $\Delta$-confluent drawings}
\label{sec:hex-grid}

Since all the internal nodes of 
underlying trees of $\Delta$-confluent graphs 
have degree three,
if uniform-length edges and 
large angular resolution 
are desirable, 
it is then natural to consider the problem of 
embedding these trees on the hexagonal grid 
where each grid point has three neighboring 
grid points and every 
cell of the grid is a regular hexagon.  

Some researchers have studied the problem of 
hexagonal grid drawing of graphs.
Kant~\cite{k-hgd-92i} presents a linear-time algorithm 
to draw tri-connected planar graphs of degree three
planar on a $n/2\times n/2$ hexagonal grid.
Aziza and~Biedl~\cite{ab-satb-05} focus on keeping 
the number of bends small.  They give algorithms 
that achieve $3.5n+3.5$ bends for all simple graphs,
prove optimal lower bounds on number of bends for $K_7$, 
and provide asymptotic lower bounds for graph classes of 
various connectivity.
We are not aware of any other result on hexagonal graph drawing, 
where the grid consists of regular hexagon cells.

\begin{figure}[htb]
  \centering
  \includegraphics[width=0.5\textwidth]{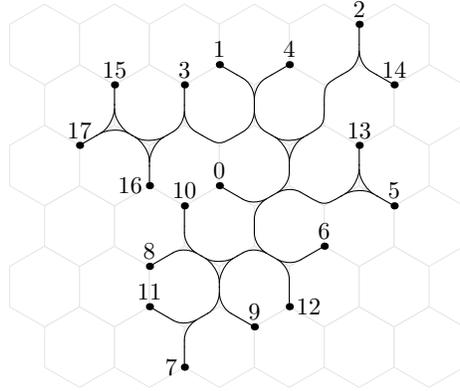}
  \caption{A hexagonal grid $\Delta$-confluent drawing example.}
  \label{fig:drawing1}
\end{figure}

In the $\Delta$-confluent drawings on the hexagonal grid, 
any segment of an edge must lie on one side of a hexagon sub-cell.
Thus the slope of any segment is
$1/2$, $\infty$, or~$-1/2$.
An example drawing for the graph from \figname~\ref{fig:dhex1}
is shown in \figname~\ref{fig:drawing1}.

Readers might notice that there are edge bends 
in the drawing of \figname~\ref{fig:drawing1}.
Some trees may require a non-constant 
number of bends per edge to be embedded on a hexagonal grid.
\journal{%
Take for example the full balanced tree,
where every internal node has degree three
and the three branches around the center of the tree have
same number of nodes.
After we choose a grid point $p$ for the center of the tree,
the cells around this point can be partitioned into layers
according to their ``distances'' to $p$ 
(\figname~\ref{fig:hex-layer}).
It is easy to see that the number of cells in layer $i$
is $6i-3$.  
That means there are $6i-3$ possible choice we can route 
the tree edges outward.
However the number of edges to children of level $i$ is 
$3\times 2^{i-1}$.
}%
Thus it is impossible to embed the tree without 
edge crossing or edge overlapping, 
when the bends are limited per edge.
\journal{%
\begin{figure}[htb]
  \centering
  \includegraphics[width=.5\textwidth]{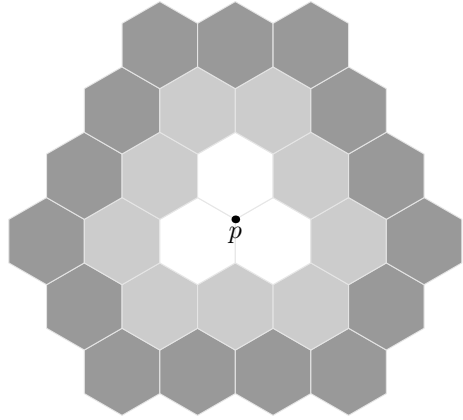}
  \caption{Layers of hexagon cells around a point.}
  \label{fig:hex-layer}
\end{figure}
}%
However, if unlimited bends are allowed, 
we show next that $\Delta$-confluent graphs can be 
embedded in the hexagonal grid of $O(n\log n)$ area in 
linear time.
%
%

\journal{%
If unlimited bends are allowed along the drawing of 
each edge, it is possible to embed 
any underlying tree of a $\Delta$-confluent graph 
on the hexagonal grid. 
}
The method is to transform an orthogonal straight-line tree 
embedding into an embedding on the hexagonal grid. 
We use 
the results of 
Chan et al.~\cite{cgkt-oaars-97}
to obtain an orthogonal straight-line tree drawing.
In their paper, a simple ``recursive winding'' approach
is presented for drawing arbitrary binary trees in small area 
with good aspect ratio.  
They consider both upward and non-upward 
cases of orthogonal straight-line drawings.  
We show that an upward orthogonal straight-line drawing of any
binary tree can be easily transformed into 
a drawing of the same tree on the hexagonal grid. 

\begin{figure}[htb]
$$
{\includegraphics[width=.4\textwidth]{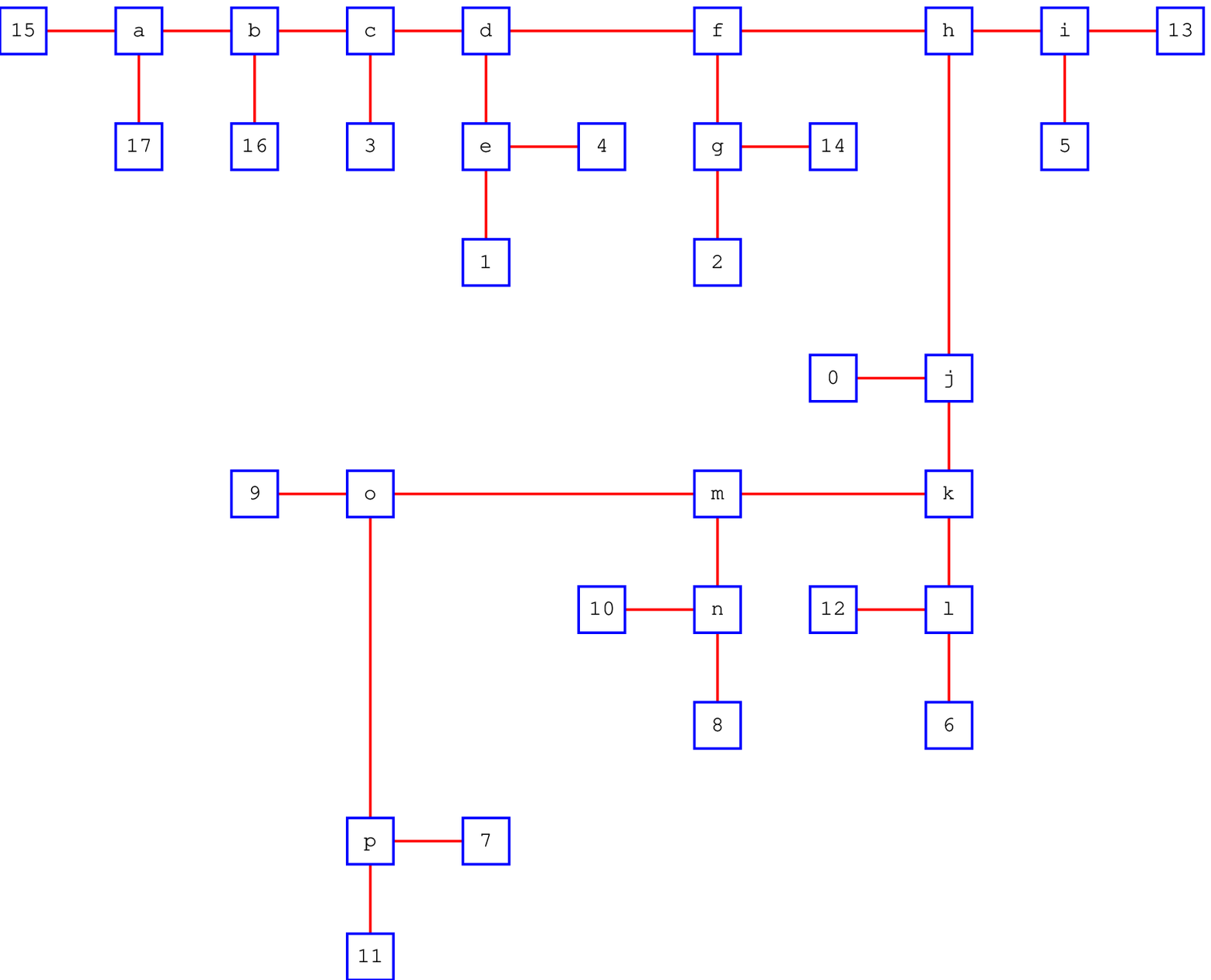}%
\atop\hbox{(a)}}\hbox{\hspace{10pt}}
{\includegraphics[width=.4\textwidth]{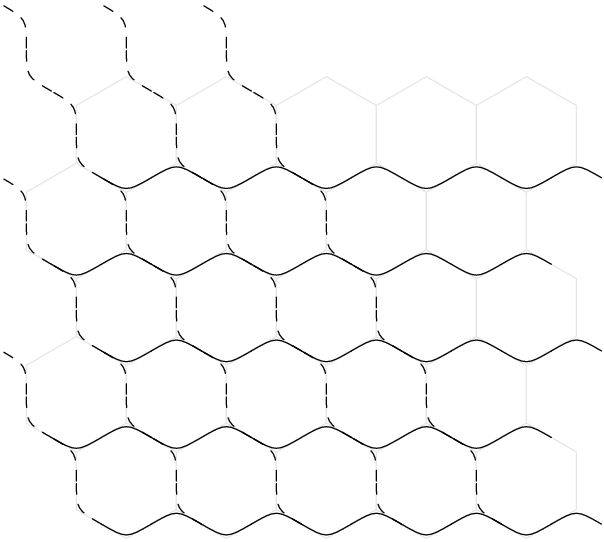}\atop\hbox{(b)}}
$$  
$$
{\includegraphics[width=.7\textwidth]{trans}\atop\hbox{(c)}}
$$
  \caption{From upward straight-line orthogonal 
    drawing to hexagonal grid drawing.  
    Internal nodes are labelled with
    letters and leaves with numbers.
    (a) orthogonal drawing, 
    generated by Graph Drawing Server (GDS)~\cite{bgt-gdtsw-97}.
    (b) $u$-curves and $v$-curves. 
    (c) unadjusted result of transformation (mirrored upside-down 
    for a change).
  }
  \label{fig:ortho2hex}
\end{figure}

\figname~\ref{fig:ortho2hex}~(a) exhibits an upward orthogonal 
straight-line drawing for the underlying tree of $G$ 
in \figname~\ref{fig:dhex1}, with node $15$ being removed 
temporarily in order to get a binary tree.

We cover the segments of the hex cell sides with two set of curves:
$u$-curves and $v$-curves (\figname~\ref{fig:ortho2hex}~$(b)$).  
The $u$-curves (solid) 
are waving horizontally and the $v$-curves (dashed) along 
one of the other two slopes.  
These two sets of curves are not direct mapping of the lines
parallel to $x$-axis or $y$-axis in an orthogonal straight-line
drawing settings, because the intersection between a $u$-curve 
and a $v$-curve is not a grid point,
but a side of the grid cell and it contains 
two grid points.  
However this does not matter very much.  
We choose the lower one of the two grid points in 
the intersection (overlapping) 
as our primary point and the other one as our backup point. 
So the primary point is at the bottom of a grid cell and its 
backup is above it to the left.
As we can see later, the backup points allow us to do a 
final adjustment of the node positions. 

When doing the transformation from an orthogonal straight-line 
drawing to a hexagonal grid drawing, we are using only the primary 
points.  So there is a one-to-one mapping between node positions 
in the orthogonal drawing and the hexagonal grid drawing. 
However, there are edges overlapping each other in the resultant 
hexagonal grid drawing of such a direct transformation
(e.g. edge $(a,b)$ and edge $(a,16)$ 
in~\figname~\ref{fig:ortho2hex}~(c)).  
Now the backup points are used to remove those overlapping portion
of edges.  Just move a node from a primary point to the point's backup 
when overlapping happens.

\begin{figure}[htb]
  \centering
  \includegraphics[width=.7\textwidth]{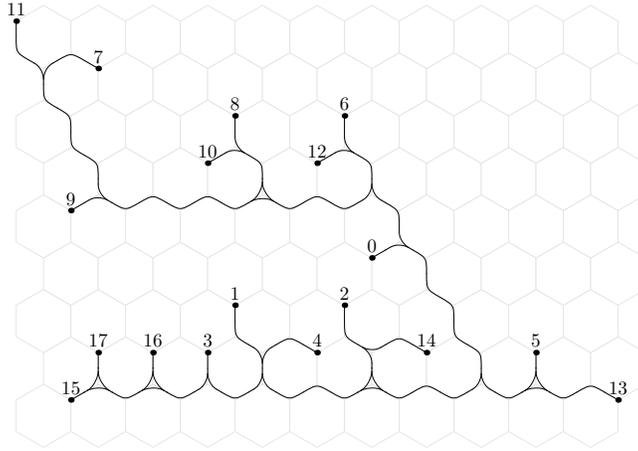}  
  \caption{Final drawing after adjustment.}
  \label{fig:final}
\end{figure}

\figname~\ref{fig:final} shows the $\Delta$-confluent drawing of $G$ 
after overlapping is removed.  
The drawing does not look compact because 
the orthogonal drawing from which it is obtained 
is not tidy in order to have the subtree separation property.

It is not hard to see that backup points are enough for removing 
all the overlapping portions while the tree structure
is still maintained.
If wanted, the backup points can be also used to reduce 
the bends along 
the edges connecting the tree leaves (e.g. edge connecting node $1$).
Some bends can be removed as well after junctions are moved 
(e.g. the subtree of node $8$ and~$10$).

\begin{thm}
Any $\Delta$-confluent graph can be embedded on
a grid of size $O(n\log n)$. 
The representation of its $\Delta$-confluent drawing 
can be computed in linear time and can be stored using
linear space.
\label{thm:thm1}
\end{thm}

\emph{Proof.} 
First the underlying tree of a $\Delta$-confluent graph 
can be computed in linear time.  
The transformation runs in linear time as well. 
It then remains to show that the 
orthogonal tree drawing can be obtained in linear time.
Chan et al.~\cite{cgkt-oaars-97}
can realize a upward orthogonal straight-line grid drawing 
of an arbitrary $n$-node binary tree $T$ 
with $O(n\log n)$ area and $O(1)$ aspect ratio.  The drawing 
achieves subtree separation and can be produced in $O(n)$ time.

By using the transformation, we can build a description of the
drawing in linear time,
which includes the placement of each vertex and
representation of each edge.
It is straightforward that the drawing
has an area of $O(n\log n)$ size. 
Since the edges are either along $u$-curves, or along $v$-curves,
we just need to store the two end points for each edge. 
Note that although some edge might contain 
$O(\sqrt{n\log n})$ bends
(from the ``recursive winding'' method),  constant amount of 
space is enough to describe each edge.  Thus the total space 
complexity of the representation is $O(n)$.
\qed

In the hexagonal grid drawings for trees,
the subtree separation property is retained 
if the subtree separation in hexagonal grid drawings is defined using 
$u,v$ area. 
If different methods of visualizing binary trees 
on the orthogonal grid are used, 
various time complexities, area requirements, 
and other drawing properties for the hexagonal grid $\Delta$-confluent drawing
can be derived as well. 

\journal{%
The same transformation could also be used to transform 
orthogonal planar drawings of graphs other than trees 
into drawings on the hexagonal grid.  
However, it will not work 
for non-planar drawings, 
because one edge crossing will be transformed into 
a hexagon cell side, 
and it brings the ambiguity 
of whether that side is a crossing or a confluent track.

If we are not limited on any grid and non-uniform edge lengths are allowed, 
it is then very natural to draw the underlying tree in a way such that 
every edge is a straight-line segment which has one of 
the three possible slopes.  
\figname~\ref{fig:flake} shows the $\Delta$-confluent drawing of a 
clique of size $3\times 2^{5}=96$.

\begin{figure}[htb]
  \centering
  \includegraphics[width=0.7\textwidth]{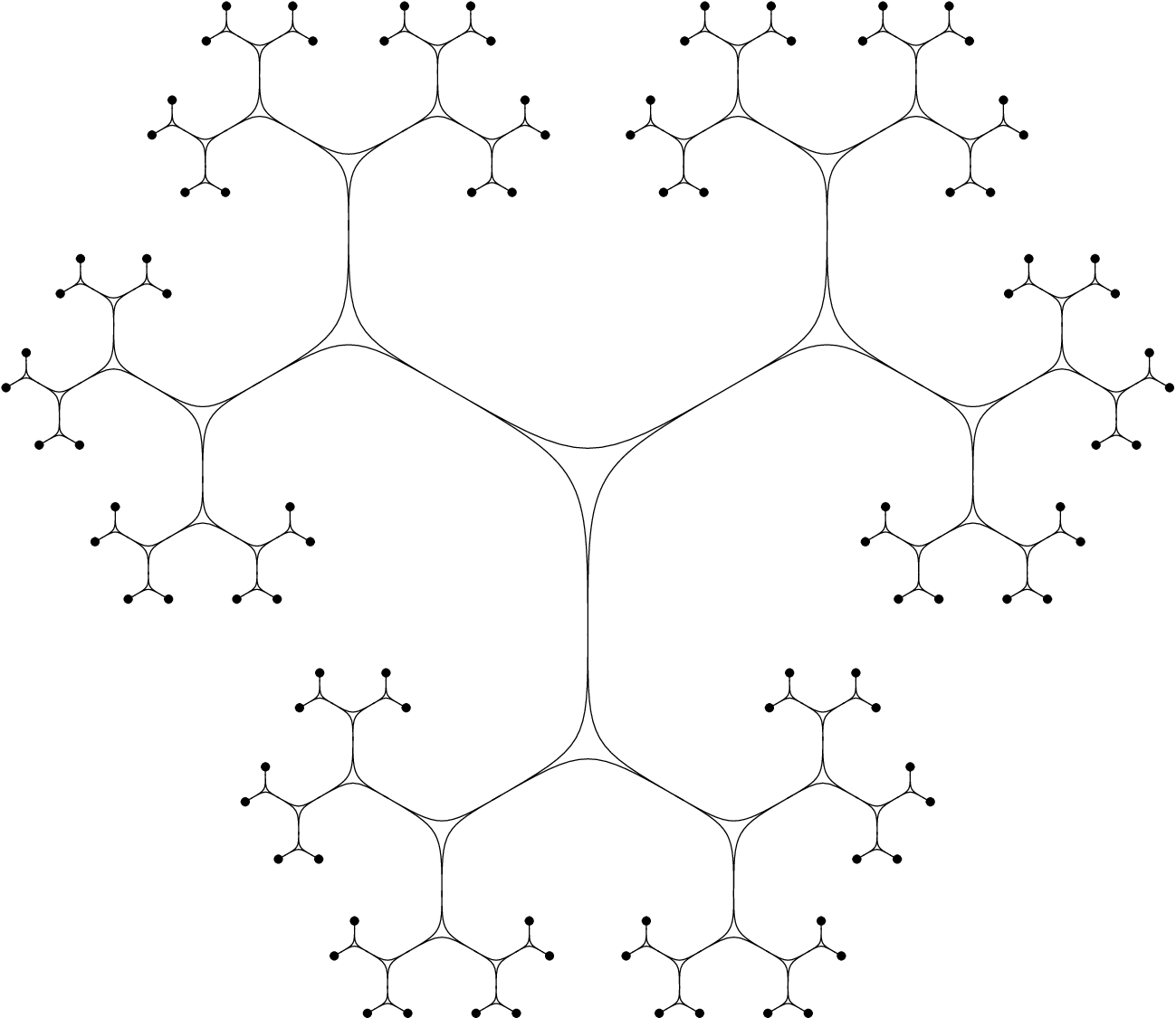}
  \caption{$\Delta$-confluent drawing of $K_{96}$.}
  \label{fig:flake}
\end{figure}

When the size of the clique is very large ($\rightarrow\infty$), 
the edge lengths of the underlying tree 
must be chosen carefully, 
otherwise the subtrees could overlap with each other,
hence introducing crossings. 
Simple calculations show that if the edge length is shortened by a same 
constant ratio (multiplied by a real number between $0$ and $1$) 
while the tree depth is increased by $1$, to avoid subtree overlapping, 
the ratio must be less than $(\sqrt{3}\sqrt{4\;\sqrt{3}+1}-\sqrt{3})/6$.

Although all junctions in \figname~\ref{fig:flake} are $\Delta$-junctions
and the underlying tree is fully balanced,  
it is easy to see that 
the same method can be used to draw $\Delta$-confluent graphs 
whose drawings have unbalanced underlying trees and
have both $\Delta$- and $\Lambda$-junctions.
}

\section{More about $\Delta$-confluent graphs}
\label{sec:more-about-delta}

In this section we discuss a $\Delta$-confluent subgraph problem, 
and list some topics of possible future work 
about $\Delta$-confluent graphs.

One way to visualize a non-planar graph is 
to find a maximum planar subgraph of the original graph, 
compute a planar drawing of the subgraph,
and add the rest of the original graph back on the drawing.
An analogous method to visualize a non-$\Delta$-confluent graph 
would be to find a maximum $\Delta$-confluent subgraph,
compute a $\Delta$-confluent drawing,
and add the rest back.  
However, just like the maximum planar subgraph problem, 
the maximum $\Delta$-confluent subgraph problem is difficult.
The problem is defined below,
and its complexity is given in Theorem~\ref{thm:npc}.

\begin{minipage}[l]{.9\linewidth}
  \vspace{.1in}
  \textsc{Maximum $\Delta$-confluent Subgraph Problem}:

  \textsc{Instance}: A graph $G = (V,E)$, an integer $K\le|V|$.
  
  \textsc{Question}: Is there a $V'\subset V$ with $|V'|\ge K$
  such that the subgraph of $G$ induced by $V'$ is a $\Delta$-confluent?
  \vspace{.1in}
\end{minipage}

\begin{thm}
Maximum $\Delta$-confluent subgraph problem is NP-complete.  
\label{thm:npc}
\end{thm}
\emph{Proof.} 
The proof can be derived easily 
from Garey and Johnson~\cite[][GT21]{gj-cigtn-79}. 

\begin{minipage}[c]{.9\linewidth}
  \vspace{.1in}
  [GT21] \textsc{Induced Subgraph with Property $\Pi$}:
  
  \textsc{Instance}:  A graph $G = (V,E)$, an integer $K \le |V|$.
  
  \textsc{Question}:   Is there a $V' \subset V$ with $|V'| \ge K$ 
  such that the subgraph of $G$ induced by $V'$ has property $\Pi$?  
  \vspace{.1in}
\end{minipage}

It is NP-hard for any property $\Pi$ that holds for arbitrarily large
graphs, does not hold for all graphs, and is hereditary (holds for all
induced subgraphs of~$G$ whenever it holds for~$G$).  If it can be
determined in polynomial time whether $\Pi$ holds for a graph, then the
problem is NP-complete.  
Examples include 
``$G$~is a clique'', 
``$G$~is an independent set'', 
``$G$~is planar'', 
``$G$~is bipartite'', 
``$G$~is chordal.''

$\Delta$-confluency is a property that holds for arbitrarily large
graphs, does not holds for all graphs, and is hereditary (every induced 
subgraph of a $\Delta$-confluent graph is $\Delta$-confluent.) 
It can be determined in linear time whether a graph is $\Delta$-confluent.
Thus the maximum $\Delta$-confluent subgraph problem 
is NP-complete. \qed

Instead of drawing the maximum subgraph $\Delta$-confluently and 
adding the rest back,  
We could compute a $\Delta$-confluent subgraph cover of the input 
graph, visualize each subgraph as a $\Delta$-confluent drawing, 
and overlay them together.  This leads to the
\textsc{$\Delta$-confluent Subgraph Covering Problem}. 
Like the 
maximum $\Delta$-confluent subgraph problem, 
we expect this problem to be 
hard as well.

This alternative way is related to 
the concept of \emph{simultaneous embedding}
(see~\cite{ek-sepgi-03,bcdeeiklm-sge-03,dek-gtldg-04,ek-sepgf-05}).
To visualize an overlay of $\Delta$-confluent subgraph drawings 
is to draw trees simultaneously. 
However \emph{simultaneously embedding} 
draws only two graphs that share the same vertex set $V$, 
while a $\Delta$-confluent subgraph cover could have 
a cardinality larger than two.  
Furthermore, the problem of simultaneously embedding 
(two) trees hasn't been solved.

Other interesting problems include:
\begin{itemize}
\item 
How to compute the drawing with optimum area 
(or number of bends, etc.) for a $\Delta$-confluent graph?

Generally hexagonal grid drawings by transforming orthogonal drawings 
are not area (number of bends, etc.) optimal.  
If subtree separation is not required, hexagonal grid drawings with 
more compact area or smaller number of bends can be achieved. 
Maybe a simple incremental algorithm would work.

\item 
The underlying track system here is topologically a tree.  
What classes of graphs can we get if other structures are allowed?

\end{itemize}

\bibliographystyle{abbrv}
\bibliography{references}

\end{document}